\documentclass[fleqn,usenatbib]{mnras}

\usepackage{newtxtext,newtxmath}

\usepackage{amsmath}
\usepackage{amssymb}
\usepackage{placeins}
\usepackage{lettrine}

\usepackage{xcolor}

\usepackage{subcaption}
\usepackage{caption}

\usepackage{tikz}
\usetikzlibrary{positioning,fit,decorations.pathreplacing,shapes.multipart,shapes.misc}

\usepackage[T1]{fontenc}

\DeclareRobustCommand{\VAN}[3]{#2}
\let\VANthebibliography\thebibliography
\def\thebibliography{\DeclareRobustCommand{\VAN}[3]{##3}\VANthebibliography}

\usepackage{graphicx}	
\usepackage{amsmath}	

\newcommand{\KT}{\href{\#cite.Krachmalnicoff2019}{KT19}}
\newcommand{\OL}{\href{\#cite.OaydaInPrep}{OL26}}
\newcommand{\BE}{\href{\#cite.Bevins2024}{B25}}
\newcommand{\LA}{\href{\#cite.10.1093/mnras/staf1621}{L25}}

\newcommand{\red}[1]{\textcolor{black}{#1}}
\newcommand{\revision}[1]{\textcolor{black}{#1}}
\newcommand{\newrevision}[1]{\textcolor{black}{#1}}

\title[\revision{Simulation-based tension quantification}]{\revision{Simulation-based tension quantification of the cosmic dipole}}

\author[M. Land-Strykowski et al.]{
Mali Land-Strykowski,$^{1,2}$\thanks{E-mail: mali.land-strykowski@sydney.edu.au (MLS)}
Harry T. J. Bevins,$^{3,4}$
Oliver T. Oayda,$^{1,2}$
and Geraint F. Lewis$^{1}$
\\
$^{1}$Sydney Institute for Astronomy, School of Physics, A28, The University of Sydney, NSW 2006, Australia\\
$^{2}$CSIRO Space and Astronomy, PO Box 76, Epping, NSW, 1710, Australia\\
$^{3}$Astrophysics Group, Cavendish Laboratory, Cambridge, CB3 0HE, UK\\
$^{4}$Kavli Institute for Cosmology, Cambridge, CB3 0HA, UK
}

\date{Accepted XXX. Received YYY; in original form ZZZ}

\pubyear{\the\year{}}

\begin{document}
\label{firstpage}
\pagerange{\pageref{firstpage}--\pageref{lastpage}}
\maketitle

\begin{abstract}
The cosmic dipole measured in surveys of cosmologically distant sources consistently exceeds the expectation \red{derived from} the cosmic microwave background, posing a significant challenge to the standard $\Lambda$CDM \red{cosmology}. In the era of precision cosmology, quantifying robust tensions is constrained by our ability to model complex, non-linear \revision{effects} that often result in intractable likelihood functions. In this paper, we present a flexible simulation-based inference \revision{(SBI)} \red{architecture} for \revision{measuring} the cosmic dipole tension using Neural Ratio Estimators (NREs). We design and train an ensemble of NREs to evaluate the \red{log} \revision{Bayesian evidence} ratio \revision{and} measure $N\sigma$ tension. We validate our approach against nested sampling, demonstrating that it accurately recovers the ground-truth. \revision{Under the kinematic interpretation of the dipole, we apply our} \red{approach} to \textit{Planck}, \revision{the} \red{National Radio Astronomy Observatory Very Large Array} Sky Survey \red{(NVSS)}, \revision{the} Rapid \revision{ASKAP} Continuum Survey \red{(RACS)}, and \revision{the} Wide-field Infrared Survey Explorer \revision{catalogue} (CatWISE). \revision{Here, leveraging SBI,} we measure a \revision{$\approx5.7\sigma$} tension between \revision{CatWISE and \textit{Planck}. We demonstrate the extensibility of our approach by applying it to forward-modelled simulations of the CatWISE Eddington bias, revealing a $\approx6.7\sigma$ tension. The methodology proposed here enables robust tension quantification as we enter the era of LSST, \textit{Euclid}, and the SKA.}
\end{abstract}




\section{Introduction}
Observations of the cosmic dipole in surveys of cosmologically distant sources are consistently at odds with the expectation set by the cosmic microwave background (CMB). This discrepancy has been measured at a significance exceeding $5\sigma$ in both conditional frequentist \citep[e.g.][]{Secrest2022} and Bayesian \revision{frameworks} \citep[e.g.][\red{and \citealt{10.1093/mnras/staf1621}, hereafter, \LA}]{10.1093/mnras/stad2322,6z32-3zf4}. Such a persistent tension poses a fundamental challenge to the Cosmological Principle and the standard concordance model of cosmology, $\Lambda$CDM \citep[for recent reviews, see][]{PEEBLES2022169159,2022JHEAp..34...49A,KumarAluri_2023,2025NatRP...7...68S,Secrest2025}.

In the decades since the proposal of the \citet{Ellis1984} test, the field has transitioned from a data-limited regime to the era of precision cosmology. Our primary constraints are no longer the statistical power of \revision{the data}, but rather our understanding of the source counts—specifically the underlying \revision{processes affecting them, such as latent survey systematics and potentially unknown astrophysical phenomena.}
\revision{While assertions by \citet{Abghari_2024} regarding systematics affecting the Wide-field Infrared Survey Explorer catalogue \citep[CatWISE;][]{2021ApJS..253....8M} have recently been refuted by \citet{10.1093/mnras/stag201}, there remains evidence of unaccounted systematics in the Rapid Australian Square Kilometre Array Pathfinder Continuum Survey \citep[RACS-low;][]{McConnell2020}, as shown by the tension analysis in \LA. Given that RACS-low has been employed in many previous works \citep[see, e.g.][\LA]{Darling_2022,10.1093/mnras/stad2161,10.1093/mnras/stae414,Wagenveld_2023,Oayda2024,2025A&A...697A.112W,10.1093/mnras/stad2322,6z32-3zf4}, the investigation of systematics remains a priority.}
\revision{Setting systematics aside, various astrophysical effects may be at play. These include large-scale inhomogeneities and local voids \citep[see, e.g.][]{Rubart2014,2025PhRvD.111j3502C}, anisotropic or tilted cosmologies \citep[see, e.g.][]{kvvs-97ly,Tzartinoglou2024}, and large-scale bulk flows \citep[see, e.g.][]{10.1093/mnras/stad1984}. Disentangling the specific contribution of each of these effects on the observed dipole tension is an active area of investigation.}

\revision{The path forward lies in incorporating both the effects of systematics and novel astrophysical phenomena into the inference pipeline.} Accounting for such effects\revision{, however,} traditionally requires understanding their impact on the expected source counts and\revision{, by extension, the form of the} likelihood. \revision{In many cases it is not possible to express this form analytically.} When analytical likelihoods are unavailable, one can simulate the \revision{desired} effects \revision{via forward models}. We therefore propose \revision{quantifying tension using} simulation-based inference (SBI), a powerful tool that allows neural networks trained on simulated data to perform Bayesian inference \red{\citep{doi:10.1073/pnas.1912789117}}.

\red{In this paper, we employ the \textsc{tensionnet} formalism introduced by \citet[][\red{hereafter, \BE}]{Bevins2024} using Neural Ratio Estimators (NREs) and adapt it to \revision{quantify} the tension between data sets \revision{used to measure the cosmic dipole}. In place of the NREs trained on summary statistics in \BE, we build our \textsc{tensionnet} from an ensemble of NREs that use a novel architecture to perform inference using \textsc{Healpix}\footnote{\url{https://healpix.sf.net}} images of the sky. For sets of experiments, we use the \textsc{tensionnet} to predict \revision{their} log \revision{Bayesian evidence} ratio \revision{(}$\log R$\revision{),} access their in concordance $\log R$ distribution, \revision{and measure} their prior-calibrated $N\sigma$ tension.}

\revision{The remainder of this work is structured as follows:} In Section~\ref{sec:background}, we outline \revision{the required} Bayesian \revision{theory, the background of NREs, and the application of them for measuring tension.} \red{We then design, train and validate our NRE architecture in Section~\ref{sec:nre-architecture}}. In Section~\ref{sec:results-baseline}, \revision{we detail our samples and simulations, and then apply the \textsc{tensionnet} to \textit{Planck} CMB observations \citep{Planck2020a}, the National Radio Astronomy Observatory Very Large Array Sky Survey \citep[NVSS;][]{Condon1998}, the Rapid Australian Square Kilometre Array Pathfinder Continuum Survey \citep[RACS-low;][]{McConnell2020} and the Wide-field Infrared Survey Explorer catalogue \citep[CatWISE;][]{2021ApJS..253....8M}. Here, under the kinematic interpretation of the dipole, we leverage SBI to present a new, prior-calibrated set of tensions.} Then, in Section~\ref{sec:results-intractable}, we \revision{showcase the utility of our architecture by applying it to a model} where the likelihood is intractable. \revision{Here, we train the \textsc{tensionnet}} on CatWISE simulations that \revision{forward-}model Eddington bias \citep[CatSIM;][\revision{hereafter, \OL}]{OaydaInPrep}. We discuss some limitations of the NRE in Section~\ref{sec:limitations} and conclude in Section~\ref{sec:conclusion}.

\section{\red{Background}}\label{sec:background}

\subsection{Bayesian theory}\label{sec:Rstat} 
Bayesian inference is a cornerstone of modern cosmological analysis\revision{; it} has been widely \revision{used to study the cosmic dipole} \citep[\revision{e.g.}][]{10.1093/mnras/stad3706,2024A&A...690A.163W,10.1093/mnras/stae2776,2025OJAp....8E.143M} \revision{and provides} a robust framework for tension quantification \citep[][]{10.1093/mnras/stab1670}. For a data set $D$ with posterior $\mathcal{P}_D$, likelihood $\mathcal{L}_D$, prior $\pi$ and evidence $\mathcal{Z}_D$, for a given set of parameters $\theta$, we use Bayes's Theorem with the notation
\begin{align}
    P(\theta|D)=\frac{P(D|\theta)P(\theta)}{P(D)}~~~\Leftrightarrow~~~\mathcal{P}_D(\theta)=\frac{\mathcal{L}_D(\theta)\pi(\theta)}{\mathcal{Z}_D}.\label{eq:bayes-theorem}
\end{align}
Here, the individual evidence for $D$ is the integral \(\mathcal{Z}_D=\int\mathcal{L}_D(\theta)\pi(\theta)d\theta\). For independent data sets $A$ and $B$, the joint evidence is \(\mathcal{Z}_{AB}=\int\mathcal{L}_{AB}(\theta)\pi(\theta)d\theta\neq\mathcal{Z}_A\mathcal{Z}_B\), where $\mathcal{L}_{AB}=\mathcal{L}_A\mathcal{L}_B$ is the joint likelihood. These individual and joint evidences form the \revision{Bayesian evidence} \red{ratio} $R$ \citep{PhysRevD.73.067302}
\begin{align}
    R=\frac{\mathcal{Z}_{AB}}{\mathcal{Z}_A\mathcal{Z}_B},\label{eq:R}
\end{align}
which quantifies the consistency between data sets. All terms in $R$ assume the same underlying model, where the data sets are explained by the same set of parameters in the numerator and each a different set of parameters in the denominator. If $R\gg1$ the data sets are consistent (greater confidence in $A$ given knowledge of $B$\red{, or vice versa}) and if $R\ll1$ the data sets are in tension (lesser confidence in $A$ given knowledge of $B$\red{, or vice versa}). This is shown \red{in} \citet{Handley2019} by interpreting $R$ through Bayes's Theorem, such that
\begin{align}
    R=\frac{P(A,B)}{P(A)P(B)}=\frac{P(A|B)}{P(A)}=\frac{P(B|A)}{P(B)}.\label{eq:R2}
\end{align}
They show that combining Equations \ref{eq:bayes-theorem} and \ref{eq:R2} \red{with the evidence integral yields}
\begin{align}
    R=\int\frac{\mathcal{P}_A(\theta)\mathcal{P}_B(\theta)}{\pi(\theta)}~d\theta=\left<\frac{\mathcal{P}_A}{\pi}\right>_{\mathcal{P}_B}=\left<\frac{\mathcal{P}_B}{\pi}\right>_{\mathcal{P}_A},\label{eq:Rpriordependence}
\end{align}
where it can be seen that $R$ is dependent on the prior volume. Here, wide priors increase $R$ and can hide tension.

The prior-dependence of $R$, however, can be calibrated-out by accessing the distribution of possible \red{$\log R$} \revision{values from experiments that are in concordance,} given \revision{a} choice of prior and model. \revision{This `}in concordance\revision{'} \red{$\log R$} distribution allows for a given value of \red{$\log R$} \red{observed} between two experiments \red{(which we denote as \red{$\log R_\text{obs}$})} to be translated into a prior-calibrated $N\sigma$ tension.

\subsection{\revision{Neural Ratio Estimators}}\label{sec:nre-background-extra}
\revision{Neural Ratio Estimators (NREs) are neural network classifiers that output the
probability that two inputs have been drawn from the same distribution as opposed to being
drawn from different distributions. They sit within the family of simulation-based
inference (SBI) methods, which replace evaluating an analytical likelihood with
sampling from a forward model \citep[for a review, see][]{doi:10.1073/pnas.1912789117}.
An optimally trained NRE learns a function of the density ratio between the two classes it is trained on \citep{cranmer2015}. Thus, if trained on an equal number of samples drawn from the joint distribution and
the product of marginal distributions, its output tends towards the ratio of those two densities.}

\revision{Crucially, the quantity an NRE learns is set entirely by the pairs of input objects it is trained on. \citet{hermans2020} first exploited this by training a classifier to separate
parameter--data pairs $(\theta,x)$ drawn from the joint $P(\theta,x)$ from those drawn from the
product of marginals $P(\theta)P(x)$. Here, the network output converges to the
likelihood-to-evidence ratio $P(x|\theta)/P(x)$ and thereby enables amortised posterior inference
with a learned likelihood \citep[see also][]{miller2021,cole2022}. If the inputs are instead labelled by which of two competing models generated them, the same machinery returns the Bayes factor between those models \citep[see, e.g.][]{jeffrey2024}. Subsequently, this has been applied to Bayesian forecasting \citep{gesseyjones2024} and experimental design optimisation \citep{leeney2026}. The application in this work follows the same principle with a different choice of input, which we explain below.}


\subsection{\revision{Measuring tension with NREs}}\label{sec:nre-background}
\revision{As mentioned, NREs learn the density ratio between the two input classes they are trained on. If we train an NRE on an equal number of examples from each class, namely data set pairs (\red{}$A$ and $B$\red{})} drawn from \red{their joint distribution (where they share the same parameters) and \revision{the product of their marginal distributions} (where they have different parameters)}, \revision{its} output tends towards
\begin{align}
    \log r = \log \frac{P(A,B)}{P(A)P(B)}.\label{eq:R-from-NRE}
\end{align}
\revision{By comparing Equations \ref{eq:R2} and \ref{eq:R-from-NRE}, we can see that $\log r = \log R$ (for a derivation see Appendix~\ref{appendix:derivation}). With the} trained NRE, \revision{we} can \revision{then} pass through \revision{pairs of} \red{simulated observations from \revision{our} experiments \revision{that share the} same parameters (\revision{they are drawn} from the joint distribution)}. \revision{These simulation pairs share the same parameters and are therefore in concordance, so passing many through the NRE evaluates} the \revision{`}in concordance\revision{'} distribution \revision{of \red{$\log R$} values}. Since the distribution is \revision{evaluated on simulations, it is inherently flexible and can capture any conceivable set of intractable effects or complex systematics that are built-in to the experiments (forward models).}

In this paper, we use the \revision{prior-calibrated tension quantification} method proposed by \BE, which we summarise \revision{below}:
\begin{enumerate}
    \renewcommand{\labelenumi}{\roman{enumi}.}
    \item Firstly, using the same prior and model that was used to compute \red{$\log R_\text{obs}$}, generate a set of matched simulations \red{of two observables} $s=\{A(\theta_i),B(\theta_i)\}^N_{i=0}$, where they share the same $\theta$, and label these with a value of one.
    \item Shuffle half of the matched simulations to construct the mismatched simulations $s'=\{A(\theta_i),B(\theta_j)\}^N_{i\neq j=0}$ and label these with a value of zero.
    \item Shuffle $s$ and $s'$ and split them into training and validation sets used to train the NRE.
    \item Generate a new set of matched simulations $z=\{A(\theta_i),B(\theta_i)\}^N_{i=0}$ and pass them through the trained NRE to evaluate the in concordance \red{$\log R$} distribution.
    \item \red{Finally, translate \red{$\log R_\text{obs}$} to tension \revision{$T$ in units of $N\sigma$\footnote{\revision{Unlike the frequentist $N\sigma$ of a null-hypothesis $p$-value, we stress that $T$ is the prior-invariant, parametrisation-invariant tension carried by the Bayesian evidence ratio $\log R$, calibrated against its in concordance distribution (see \BE).}}} using} the cumulative distribution function (CDF) of the in concordance \red{$\log R$} distribution\red{, $\alpha=P(X<X')$ where $X=\log R$,} and the \red{one-sided} inverse survival function of the standard normal distribution
    \begin{align}
       \red{T=z\left(\frac{\alpha}{2}\right)=\sqrt2~\text{\red{e}rf}^{-1}(1-\alpha),\label{eq:T}}
    \end{align}
    \red{where} \red{e}rf$^{-1}$ is the inverse error function. \red{The prior-calibrated $N\sigma$ tension, $T$,} informs us of the relationship between our experiments, given our prior and model. Here, $T=3$ indicates that our experiments are in $3\sigma$ tension, and $T=0$ corresponds to perfect agreement between our experiments.
\end{enumerate}


\section{\red{Method}}\label{sec:nre-architecture}

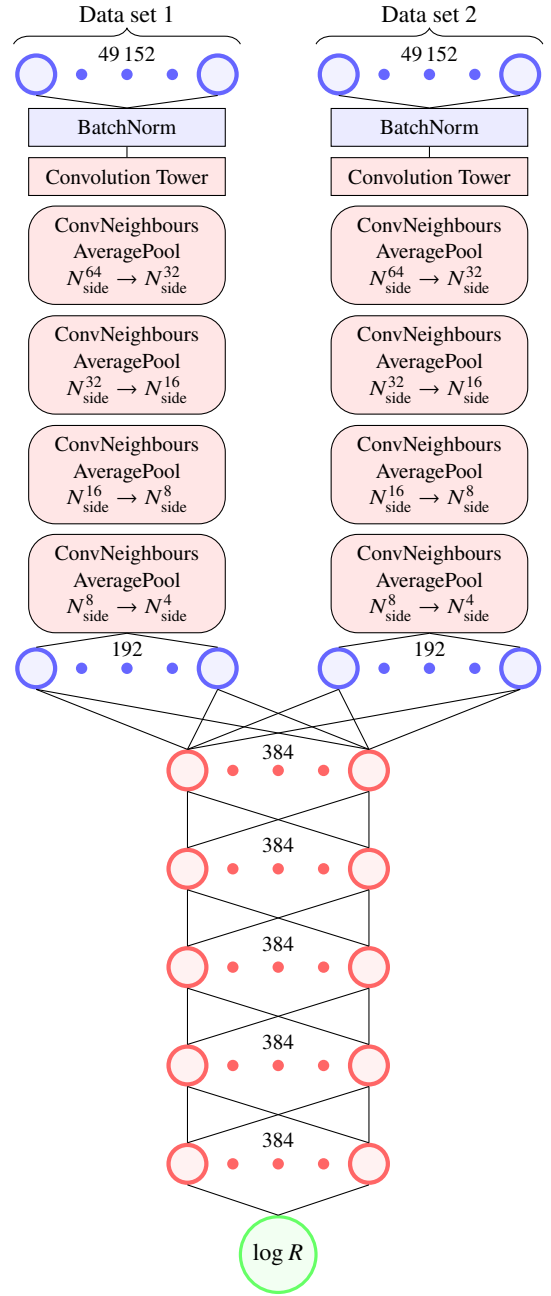
\begin{figure}
    \centering
    \begin{tikzpicture}[
  rednode/.style={circle, draw=red!60, fill=red!5, ultra thick, minimum size=5mm},
  bluenode/.style={circle, draw=blue!60, fill=blue!5, ultra thick, minimum size=5mm},
  greennode/.style={circle, draw=green!60, fill=green!5, very thick, minimum size=10mm},
  small-dot/.style args={#1}{circle, fill=#1, minimum size=1.5mm, inner sep=0pt},
  layerbox/.style={rectangle, draw=black, rounded corners=3pt, minimum width=3.6cm, align=center, inner sep=3pt},
  convbox/.style={rectangle, draw=black, fill=red!10, rounded corners=10pt, minimum width=2.6cm, minimum height=1.3cm, align=center, inner sep=0pt},
  batchbox/.style={rectangle, draw=black, fill=blue!8, rounded corners=0pt, minimum width=2.6cm, minimum height=0.5cm, align=center, inner sep=4pt},
  pinkbox/.style={rectangle, draw=black, fill=red!10, rounded corners=0pt, minimum width=2.6cm, minimum height=0.5cm, align=center, inner sep=3pt},
  every node/.style={font=\small},
  >=stealth
  ]

    \node[font=\normalsize] (ds1) at (-3,0) {Data set 1};
    \node[font=\normalsize] (ds2) at (1,0)  {Data set 2};
    
    \draw[decorate, decoration={brace, amplitude=6pt}] (-4.5,-0.4) -- (-1.5,-0.4) node[midway, yshift=-10pt,font=\footnotesize] {}; 
    \draw[decorate, decoration={brace, amplitude=6pt}] (-0.5,-0.4)  -- (2.5,-0.4)  node[midway, yshift=-10pt,font=\footnotesize] {}; 
    
    \node[bluenode] (L_top_a) at (-4.2,-0.8) {};
    \node[small-dot={blue!60}] (L_dot1)  at (-3.6,-0.8) {};
    \node[small-dot={blue!60}] (L_dot2)  at (-3,-0.8) {};
    \node[small-dot={blue!60}] (L_dot3)  at (-2.4,-0.8) {};
    \node[bluenode] (L_top_b) at (-1.8,-0.8) {};
    \node[font=\footnotesize] at (-3,-0.53) {49\,152};
    
    \node[batchbox] (L_batch) at (-3,-1.5) {BatchNorm};
    \draw (L_top_a.south) -- (L_batch.north);
    \draw (L_top_b.south) -- (L_batch.north);
    
    \node[pinkbox] (L_convtower) at (-3,-2.15) {Convolution Tower};
    \draw (L_batch.south) -- (L_convtower.north);
    
    \node[convbox] (L_conv1) at (-3,-3.19) {ConvNeighbours \\ AveragePool \\ $N_{\mathrm{side}}^{64}\to N_{\mathrm{side}}^{32}$};
    \node[convbox] (L_conv2) at (-3,-4.64) {ConvNeighbours \\ AveragePool \\ $N_{\mathrm{side}}^{32}\to N_{\mathrm{side}}^{16}$};
    \node[convbox] (L_conv3) at (-3,-6.09) {ConvNeighbours \\ AveragePool \\ $N_{\mathrm{side}}^{16}\to N_{\mathrm{side}}^{8}$};
    \node[convbox] (L_conv4) at (-3,-7.53) {ConvNeighbours \\ AveragePool \\ $N_{\mathrm{side}}^{8}\to N_{\mathrm{side}}^{4}$};
    
    \node[bluenode] (L_bot_a) at (-4.2,-8.65) {};
    \node[small-dot={blue!60}] (L_bd1)  at (-3.6,-8.65) {};
    \node[small-dot={blue!60}] (L_bd2)  at (-3,-8.65) {};
    \node[small-dot={blue!60}] (L_bd3)  at (-2.4,-8.65) {};
    \node[bluenode] (L_bot_b) at (-1.8,-8.65) {};
    \node[font=\footnotesize] at (-3,-8.4) {192};
    
    \draw (L_conv4.south) -- (L_bot_a.north);
    \draw (L_conv4.south) -- (L_bot_b.north);
    
    \node[bluenode] (R_top_a) at (-0.2,-0.8) {};
    \node[small-dot={blue!60}] (R_dot1)  at (0.4,-0.8) {};
    \node[small-dot={blue!60}] (R_dot2)  at (1,-0.8) {};
    \node[small-dot={blue!60}] (R_dot3)  at (1.6,-0.8) {};
    \node[bluenode] (R_top_b) at (2.2,-0.8) {};
    \node[font=\footnotesize] at (1,-0.53) {49\,152};
    
    \node[batchbox] (R_batch) at (1,-1.5) {BatchNorm};
    \draw (R_top_a.south) -- (R_batch.north);
    \draw (R_top_b.south) -- (R_batch.north);
    
    \node[pinkbox] (R_convtower) at (1,-2.15) {Convolution Tower};
    \draw (R_batch.south) -- (R_convtower.north);
    
    \node[convbox] (R_conv1) at (1,-3.19) {ConvNeighbours \\ AveragePool \\ $N_{\mathrm{side}}^{64}\to N_{\mathrm{side}}^{32}$};
    \node[convbox] (R_conv2) at (1,-4.64) {ConvNeighbours \\ AveragePool \\ $N_{\mathrm{side}}^{32}\to N_{\mathrm{side}}^{16}$};
    \node[convbox] (R_conv3) at (1,-6.09) {ConvNeighbours \\ AveragePool \\ $N_{\mathrm{side}}^{16}\to N_{\mathrm{side}}^{8}$};
    \node[convbox] (R_conv4) at (1,-7.53) {ConvNeighbours \\ AveragePool \\ $N_{\mathrm{side}}^{8}\to N_{\mathrm{side}}^{4}$};
    
    \node[bluenode] (R_bot_a) at (-0.2,-8.65) {};
    \node[small-dot={blue!60}] (R_bd1)  at (0.4,-8.65) {};
    \node[small-dot={blue!60}] (R_bd2)  at (1,-8.65) {};
    \node[small-dot={blue!60}] (R_bd3)  at (1.6,-8.65) {};
    \node[bluenode] (R_bot_b) at (2.2,-8.65) {};
    \node[font=\footnotesize] at (1,-8.4) {192};
    
    \draw (R_conv4.south) -- (R_bot_a.north);
    \draw (R_conv4.south) -- (R_bot_b.north);
    
    \def\fcstarty{-8.6-1.4}
    \def\fcysep{1.3} 
    
    \foreach \i in {0,...,4} {
    \pgfmathsetmacro{\yy}{\fcstarty - \i*\fcysep}
    \node[rednode] (fc\i a) at (-2.2,\yy) {};
    \node[small-dot={red!60}] (fc\i d1) at (-1.6,\yy) {};
    \node[small-dot={red!60}] (fc\i d2) at (-1.0,\yy) {};
    \node[small-dot={red!60}] (fc\i d3) at (-0.4,\yy) {};
    \node[rednode] (fc\i b) at (0.2,\yy) {};

    \ifnum\i=0
        \def\labelyshift{0.25} 
    \else
        \def\labelyshift{0.31}  
    \fi

    \node[font=\footnotesize] at ( -1.0 , \yy+\labelyshift) {384};
}
    
    \foreach \s in {L_bot_a,L_bot_b,R_bot_a,R_bot_b}{
      \foreach \t in {fc0a,fc0b}{
        \draw (\s.south) -- (\t.north);
      }
    }
    
    \foreach \i in {0,...,3}{
      \foreach \a in {a,b}{
        \foreach \b in {a,b}{
          \draw (fc\i \a.south) -- (fc\the\numexpr\i+1\relax \b.north);
        }
      }
    }
    
    \node[greennode] (out) at (-1,-\fcstarty - 5*\fcysep - 0.3 - 16.8) {}; 
    \node at (out.center) {\normalsize $\log R$};
    
    \foreach \a in {a,b}{
      \draw (fc4\a.south) -- (out.north);
    }
    
    
    \end{tikzpicture}
    \caption{Schematic of the Neural Ratio Estimator (NRE) used in this work, which we refer to as a \textsc{tensionnet}. The numbers quoted represent the number of neurons in each respective layer. Two convolution towers compress each input data set from 49\,152 pixels to 192, before feeding them into the fully connected network, where the information from the two data sets is mixed to predict $\log R$. The training and validation sets consist of matched and mismatched simulation pairs, as described in Section~\ref{sec:nre-background}. We describe the training of the NRE in Section~\ref{sec:training-nre} and we validate it in Section~\ref{sec:nre-validation}.}
    \label{fig:architecture}
\end{figure}

\red{\revision{To employ} simulation-based tension quantification \revision{for} data sets \revision{used to measure the cosmic dipole}, we adapt the \textsc{tensionnet} formalism introduced in \BE. In particular, we design a novel NRE architecture that performs inference over the \textsc{Healpix} sky. Here, we detail \revision{the} design, training, and validation \revision{of our NRE architecture}.}

\subsection{\red{Designing the NRE}}
\red{Our NRE architecture} consists of two convolution towers that compress the information of each data set into a latent space, before mixing the information in a fully connected network to compute $\log R$. \red{To infer the optimal hyperparemeters for our architecture, we used \textsc{Optuna} \citep{akiba2019optuna}, a hyperparameter optimisation framework (see Appendix~\ref{appendix:sec:2} for details).} We visualise our network architecture in Figure~\ref{fig:architecture} and explain its design below.

\subsubsection{Convolution towers}

Before passing any data to the NRE, we first bin the simulations into equal-area sky pixels with \red{$N_{\text{side}} = 64$ (shortened to $N^{64}_{\text{side}}$ for notational conciseness)} using \textsc{Healpy} \citep{2005ApJ...622..759G,Zonca2019}, a \textsc{Python} implementation of \textsc{Healpix}. We then batch-normalise each $N_{\text{side}}^{64}$ data set as we pass it into the convolution tower. To extract features from the data set during compression, we use \textsc{NNhealpix} \citep[][\red{hereafter, \KT}]{Krachmalnicoff2019} to perform convolution on the \textsc{Healpix} sphere using one filter. Here, standard 2D convolutions cannot be directly applied to the \textsc{Healpix} sphere due to the tesselation pattern of the pixels, where most pixels have eight neighbours and some have seven (see figure 2 in \KT). \textsc{NNhealpix} unravels the neighbours of each pixel into a 1D vector and applies a 1D kernel with stride nine, where the last pixel in the seven-neighbour-case is set to zero (see figure 3 in \KT). This performs a convolution over the \textsc{Healpix} sphere, producing a new convolved data set at the same $N_{\text{side}}$ for each filter (see the ConvNeighbours step in Figure~\ref{fig:architecture}).

After feature extraction, we downscale the original $N_{\text{side}}^{64}$ data set to $N_{\text{side}}^{32}$ by average-pooling each set of four $N_{\text{side}}^{64}$ sub-pixels that correspond to a given $N_{\text{side}}^{32}$ super-pixel (see figure 4 in \KT). \red{Since some observational data sets do not cover the full sky,} we use our own custom implementation of the \textsc{NNhealpix} average-pool function that allows for masked sub-pixels (see the AveragePool step in Figure~\ref{fig:architecture}). Here, we set masked sub-pixels to a characteristic value (zero) and simply do not include them in the average. We then repeat the convolution and pooling processes until the data set is downscaled to $N_{\text{side}}^{4}$, representing a compression from 49\,152 pixels to 192.

\subsubsection{Fully connected network}
Before we pass the latent variables into the fully connected network, we flatten and concatenate the extracted features and 192 pixels for each data set. The fully connected network consists of five dense layers of 384 neurons, each followed by a \textsc{ReLU} activation and a dropout layer with a rate of 20 percent. The output layer with one neuron then performs a linear activation to output $\log R$ for the data set pair.

\subsection{\red{Training the NRE}}\label{sec:training-nre}

\red{During testing, we found that larger training data sets improved the performance of the NRE. Therefore, to fully utilize the capacity of the GPU node's SSD, we} use 900\,000 simulated data set pairs with a split of 60 percent for training, 30 percent for validation, and 10 percent to build the in concordance \red{$\log R$} distribution. We use an ADAM optimiser with an exponentially decaying learning rate starting at $10^{-3}$ with a step size of 1000 and a decay factor of 0.9. We train the NRE for a maximum of 1000 epochs with a batch size of 6000 and use early stopping with a patience of 50, monitoring validation loss. We use the binary cross entropy loss function
\begin{align}
    l=-\frac{1}{N}\left[\sum^N_i y_i\log(p_i)+(1-y_i)\log(1-p_i)\right],
\end{align}
where $p=S_\sigma(\log R)$ and $S_\sigma$ is the sigmoid activation function.

To account for the stochastic nature of the training, we reinitialise the random weights and retrain the NRE four additional times, for a total of five independent NREs \red{for each pair of observational data sets}. We then use the five NREs in an ensemble when predicting the final in concordance \red{$\log R$} distribution. Here, for each $\log R$ prediction, we randomly select an NRE according to its softmax probability
$p_i = {e^{-\ell_i}}/{Z}$, where $\ell_i$ is the \red{best} validation loss score of the $i$th NRE and $Z = \sum_{i=1}^{N} {e^{-\ell_i}}$ is the normalisation constant.

\subsection{Validating the NRE}\label{sec:nre-validation}
\subsubsection{\red{Comparing the NRE to a nested sampling ground-truth}}
To investigate the performance of our NRE architecture, we compare the in concordance \red{$\log R$} distribution predicted by our NRE to a ground-truth in concordance \red{$\log R$} distribution that is accessed using nested sampling. In principle, a nested sampling ground-truth is only accessible if the likelihood is known. Therefore, for validating the \textsc{tensionnet}, we use the kinematic interpretation of the dipole, which has a known likelihood function.

We first train five NREs as described in Section~\ref{sec:training-nre}, using simulation pairs of \textit{Planck} and RACS-low based off of the samples used in \LA~and \citet{Oayda2024}, respectively. We generate the simulations by modulating a homogeneous $N_{\mathrm{side}}^{64}$ sky with a dipole and drawing Poissonian or Gaussian deviates. We cover the samples and simulations in greater detail in Section~\ref{sec:results-baseline}.

With the simulations in hand, we select 10\,000 in concordance (matched) simulation pairs and predict their $\log R$ values using the ensemble of trained NREs, labelling this in concordance \red{$\log R$} distribution as $\mathcal{P}_{\text{NRE}}(\log R)$. We then use nested sampling to compute $\log R$ for the same 10\,000 pairs, which requires evaluating their individual and joint evidences over a total of 30\,000 chains. Here, we use a likelihood of the product of the probability of observing each unmasked pixel across the set $n_\text{pix}$
\begin{align}
\mathcal{L}(D|\Theta)=\prod_{i=1}^{n_\text{pix}}P(N_i|\Theta),
\end{align}
where $N_i$ is the value at pixel $\mathbf{\hat{n}}_i$ and $\Theta$ is a set of parameter values drawn from the same priors as the simulations. \red{In line with the Poisson-point process of counting sources in pixels, for} the RACS-low simulations we use the Poissonian form
\begin{align}
P(N_i|\Theta)=\frac{\lambda_i^{N_i}e^{-\lambda_i}}{N_i!}\red{.}\label{eq:poisson}
\end{align}
\red{As motivated by the continuous temperature measurement at each pixel, for} the \textit{Planck} simulations we use the Gaussian form 
\begin{align}
P(N_i|\Theta)= \mathcal{G}_\text{PDF}(x=N_i,\mu=\lambda_i,\sigma=\sigma_D),\label{eq:gaussian}
\end{align}
where $\sigma_D$ is the standard deviation of the temperature values from our actual \textit{Planck} sample. Here, the expectation value at pixel $\mathbf{\hat{n}}_i$, at an angle of $\theta_i$ to the dipole pointing towards ($l,b$), is given to first order by the sum of the dipolar modulation
\begin{align}
\lambda_i(\Theta)&=\bar{N}(1+\mathbf{d}\cdot\mathbf{\hat{n}}_i)\label{eq:expectation-value-vector}\\
&=\bar{N}(1+\mathcal{D}\cos\theta_i),\label{eq:expectation-value}
\end{align}
where $\mathcal{D}$ is the amplitude of the dipole $\mathbf{d}$ and $\bar{N}$ is the mean value per unmasked pixel.

We measure \red{$\log R$} by evaluating the evidence using \textsc{Anesthetic}\footnote{\url{https://github.com/handley-lab/anesthetic}} \citep{anesthetic} on samples computed via the Monte Carlo algorithm \textsc{MLFriends} \citep{2016S&C....26..383B,2019PASP..131j8005B} with \textsc{UltraNest}\footnote{\url{https://johannesbuchner.github.io/UltraNest}} \citep{2021JOSS....6.3001B}. To account for the error on each \red{$\log R$} measurement, we resample it five times from a Gaussian centred at the initial \red{$\log R$} value with standard deviation equal to its $1\sigma$ error. We label the resulting ground-truth in concordance \red{$\log R$} distribution as $\mathcal{P}_{\text{NS}}(\log R)$.

\begin{figure}
    \includegraphics[width=\columnwidth]{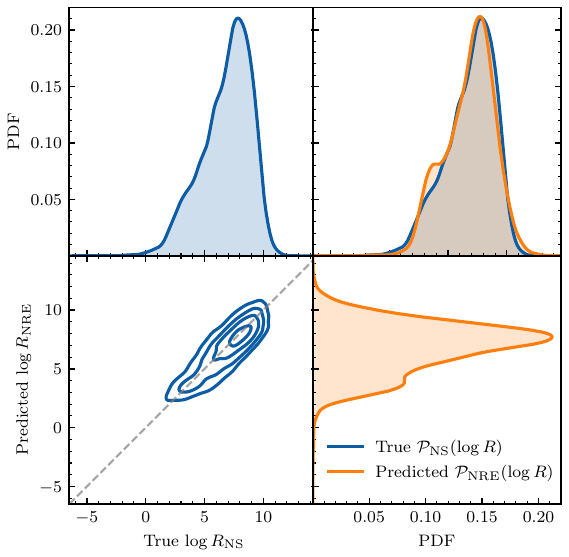}
    \caption{Comparison of the in concordance \red{$\log R$} distribution predicted by our NRE ensemble to the true distribution accessed using nested sampling. The distributions are normalised as a probability density function (PDF) and the probability density of the contours in the bottom-left panel correspond to 0.5, 1, 1.5 and 2$\sigma$. We find that $\mathcal{P}_{\text{NRE}}(\log R)$ closely matches $\mathcal{P}_{\text{NS}}(\log R)$, indicating that our \textsc{tensionnet} is performing as intended.}
    \label{fig:corner}
\end{figure}

We present the comparison of the two distributions in Figure~\ref{fig:corner}. Here, it is immediately evident that $\mathcal{P}_{\text{NRE}}(\log R)$ closely matches $\mathcal{P}_{\text{NS}}(\log R)$, since the predicted $\log R$ values cluster along the linear relation $\log R_\text{NS}=\log R_\text{NRE}$. In addition, we compute the Kullback-Leibler Divergence \citep{10.1214/aoms/1177729694}
\begin{equation}
\mathscr{D}_{\mathrm{KL}}
= \int \mathcal{P}_{\text{NS}}(\log R)
\, \log \frac{\mathcal{P}_{\text{NS}}(\log R)}{\mathcal{P}_{\text{NRE}}(\log R)}
\, \mathrm{d}\log R\label{eq:kl}
\end{equation}
and the Kolmogorov-Smirnov statistic \citep{smirnov1948}
\begin{equation}
\mathscr{D}_{\mathrm{KS}}
= \sup_{\log R}
\left|
F_{\text{NS}}(\log R)
- F_{\text{NRE}}(\log R)
\right|,
\label{eq:ks}
\end{equation}
which both quantify how well $\mathcal{P}_{\text{NRE}}(\log R)$ captures $\mathcal{P}_{\text{NS}}(\log R)$, resulting in $\log\mathscr{D}_{\mathrm{KL}}=-3.99$ and $\log\mathscr{D}_{\mathrm{KS}}=-3.53$.

\begin{figure}
    \includegraphics[width=\columnwidth]{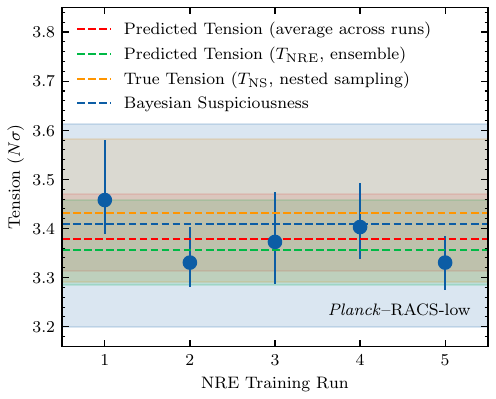}
    \caption{The resulting tension measurements between \textit{Planck} and RACS-low using the in concordance \red{$\log R$} distribution predicted by our NREs, the true distribution accessed using nested sampling, and Bayesian suspiciousness. \red{The} tension predicted \red{by} each individual NRE \red{across the training runs is represented by the blue points. The true (nested sampling), predicted (average and ensemble), and Bayesian suspiciousness tensions are represented by the dashed lines, with the shaded regions showing their respective $1\sigma$ errors.} Whilst we show the mean across these individual measurements for completeness, when using the \textsc{tensionnet} we report the ensemble result. Here, we see that the \textsc{tensionnet} successfully recovers the ground-truth, since the ensemble tension $T_\text{NRE}$ strongly agrees with the nested sampling tension $T_\text{NS}$.
    }
    \label{fig:t-comparison}
\end{figure}

We then take the actual \red{$\log R_\text{obs}$} between \textit{Planck} and RACS-low and pass it through the in concordance \red{$\log R$} distributions to compute their $N\sigma$ tension\red{, $T$ (see Equation~\ref{eq:T})}. We find that $\mathcal{P}_{\text{NS}}(\log R)$ and $\mathcal{P}_{\text{NRE}}(\log R)$ give $T_\text{NS}=3.43\substack{+0.15\\-0.14}$ and $T_\text{NRE}=3.36\substack{+0.10\\-0.07}$, which are in very strong agreement (see Figure~\ref{fig:t-comparison}). Here, both the true tension ($T_\text{NS}$) and the predicted tension ($T_\text{NRE}$) align with the Bayesian suspiciousness tension from the literature (\LA), however, that should not be expected to generally be the case. Unlike $T$, Bayesian suspiciousness \red{$N\sigma$ tension} assumes Gaussian posteriors, resulting in the two tension \red{metrics} not \red{necessarily} being equal\red{. Based on our} analysis of a toy problem comparing the two tension metrics\red{, we find that the NRE better reflects the ground-truth (see Appendix~\ref{appendix:sec:1} for details).}

The matching distributions in Figure~\ref{fig:corner} and the aligned tensions in Figure~\ref{fig:t-comparison} give us confidence in the architecture of \red{our} \textsc{tensionnet}. We proceed to apply our \textsc{tensionnet} to surveys currently used to measure the dipole in Section~\ref{sec:results-baseline} (\textit{Planck}, NVSS, RACS-low, and CatWISE) and to a novel dipole model with an intractable likelihood in Section~\ref{sec:results-intractable} (CatSIM).

\subsubsection{\red{Performance comparison to nested sampling}}
\red{In reference to nested sampling, we compare our NRE to that in \BE, which they applied to Baryon Acoustic Oscillation observations and toy 21-cm problems. They reported that generating simulations for their NRE, training it, and using it to evaluate an in concordance $\log R$ distribution of 5000 pairs was 152 times faster than using nested sampling. We find that our NRE takes 10 hours to train using four NVIDIA Tesla Volta V100-SXM2-32~GB GPUs\footnote{Due to the \red{Portable Batch System (PBS)} queue structure, the GPUs are requested alongside 48 Intel Xeon Platinum 8268 (2.9~GHz) cores. However, these are not used to train the NRE.}. In comparison, using nested sampling to compute the same in concordance \red{$\log R$} distribution would take about 167 hours on 48 Intel Xeon Platinum 8274 (3.2~GHz) cores. Whilst our NRE is about 17 times faster than nested sampling, it falls short of the improvement reported in \BE, which we attribute to the complexity of the problem our NRE is tasked with. Here, for processing \textsc{Healpix} skies, our NRE uses convolutional layers, which we found were required after testing different architectures.} 

\section{Applying the NRE to Planck, NVSS, RACS-low and CatWISE}\label{sec:results-baseline}
With the \textsc{tensionnet} designed and tested for \textit{Planck} and RACS-low, we proceed to apply it to all data set combinations of \textit{Planck}, NVSS, RACS-low and CatWISE. We first detail our samples and simulations, and then we present our results and discuss them.

\subsection{Samples}
For training the NRE and building the in concordance \red{$\log R$} distribution, we use samples and simulations binned into $N_\text{side}^{64}$ \textsc{Healpixels} using \textsc{Healpy}. Our samples match those used in \cite{Secrest2022}, \citet{Oayda2024}, and \LA~\red{(see figures 1 and 2 in \LA)}.

\subsubsection{\textit{Planck}}
After \red{its discovery in the 60s, the CMB} has been observed by the Cosmic Background Explorer \citep[\textit{COBE};][]{1982OptEn..21..769M,1992ApJ...397..420B}, the Wilkinson Microwave Anisotropy Probe \citep[\textit{WMAP};][]{2003ApJ...583....1B} and the European Space Agency's \textit{Planck}\footnote{\url{https://www.esa.int/Planck}} satellites, \red{each} attaining increasingly precise measurements of our heliocentric motion derived from the dipole\red{. The canonical value today is} $v_\text{CMB}\!=\!369.82\pm0.11$~km~$\text{s}^{-1}$ towards $(l,b)\!=\!(264.021\pm0.011^\circ, 48.253\pm0.005^\circ)$\red{, as measured in the final \textit{Planck} 2018 data release} \citep{2020A&A...641A...1P,2020A&A...641A...3P}. In this work, we use the \textsc{BeyondPlanck}\footnote{\url{https://beyondplanck.science}} Data Release II \citep{2023A&A...675A...1B} of the CMB as observed by \textit{Planck} at 30~GHz using the Low Frequency Instrument \citep[LFI;][]{2010A&A...520A...4B,2011A&A...536A...3M}. We construct our sample by removing the foreground galaxy component using the SMICA intensity map \citep{2008ISTSP...2..735C} from the \textit{Planck} Collaboration Public Data Release~3 \citep[PR3;][]{2020ipac.data.I558P}.

\subsubsection{NVSS}
NVSS is a 1.4\,GHz continuum survey of 1.8 million radio sources observed between 1993 and 1997, covering the entire sky north of declination $\delta \ge -40^\circ$ \citep{Condon1998}. NVSS has been widely used in dipole analyses, with the first measurement by \citet{2002Natur.416..150B} recovering a dipole that aligns with the CMB direction but with an amplitude exceeding the expectation by a factor of 1.5--2. In the years following, many studies have measured a dipole in NVSS exceeding two \citep[e.g.][]{Singal_2011,10.1111/j.1365-2966.2012.22032.x,Rubart_2013,10.1093/mnras/stu2535,Tiwari_2015,10.1093/mnras/stx1631} and three \citep[e.g.][\LA]{Secrest2022,Wagenveld_2023,Oayda2024,2025A&A...697A.112W} times the expected kinematic amplitude. In this work, we use the NVSS `B' sample from \citet{Oayda2024} of 338\,222 sources with a flux density distribution of $15~\text{mJy}\leq S\leq1000~\text{mJy}$. The mask covers localised areas with spuriously high source counts, the Galactic plane ($|b|\leq10^{\circ}$), and an additional degree above the survey declination limit ($-40^{\circ}\leq\delta\leq-41^{\circ}$). Relatively local ($z<0.01$) sources have been cross-matched and removed using the Two Micron All Sky Survey Redshift Survey \citep[2MRS;][]{Huchra_2012} and NASA/IPAC Extragalactic Database (NED)\footnote{\url{https://ned.ipac.caltech.edu}}.

\subsubsection{RACS-low}
RACS-low (specifically\red{, its first epoch,} RACS-low1) is a large-area survey centred at 887.5\,MHz conducted between 2019 and 2020, containing about 2.1 million radio sources in the region $-80^\circ \le \delta \le 30^\circ$ and $|b| > 5^\circ$ \citep{McConnell2020,Hale2021}. With the exception of an initial measurement that combined the RACS-low data with the Very Large Array Sky Survey \citep[VLASS;][]{Lacy_2020} at $\delta=0^\circ$ by scaling the RACS-low fluxes at 887.5~MHz to those of VLASS at 3~GHz \citep{Darling_2022}, every dipole measurement with RACS-low has recovered amplitudes that are three or more times the kinematic expectation \citep[e.g.][\LA]{10.1093/mnras/stad2161,10.1093/mnras/stae414,Wagenveld_2023,Oayda2024,2025A&A...697A.112W}. In this work, we use the RACS-low `B' sample from \citet{Oayda2024} of 459\,276 sources with a flux density distribution of $15~\text{mJy}\leq S\leq1000~\text{mJy}$. The mask covers localised areas with spuriously \red{low} source counts, pixels at the edges of the Galactic plane ($|b|\!\leq\!5^\circ$) that have been partially masked, a $13^\circ$ disc around the southern equatorial pole, and an additional degree above the survey declination limit ($-77^{\circ}\leq\delta\leq-41^{\circ}$). Relatively local sources have been removed, similarly to NVSS `B'.

\subsubsection{CatWISE}
The CatWISE2020 catalogue \citep{2021ApJS..253....8M} contains 1.9 billion sources derived from all-sky 3.4 and 4.6~\textmu m (W1 and W2) observations from NASA's \textit{WISE} spacecraft \citep{Wright_2010} between 2010 and 2018. The dipoles measured with CatWISE all exceed two times the expected kinematic amplitude, being confirmed in both conditional frequentist \citep{Secrest_2021,Secrest2022} and Bayesian \citep[][\LA, \OL]{10.1093/mnras/stad2322,2025A&A...697A.112W} frameworks \citep[see also][]{Abghari_2024,bashir2025,10.1093/mnras/stag201}. Whilst \citet{Abghari_2024} suggested that the quadrupole mode in the CatWISE mask can couple the dipole with an octupole and reduce the significance of the dipole \red{amplitude} reported in previous studies, this has been recently refuted by \citet{10.1093/mnras/stag201}. In this work, we employ the sample of 1\,621\,329 infrared-selected quasars first used in \citet{Secrest2022}, with a magnitude cut of $9<\text{W1}<16.5$ and a colour cut of $\text{W1} - \text{W2} > 0.8$ \citep{Stern_2012}. The mask covers known bright sources, the Galactic plane ($|b|<30^\circ$), and localised regions with poor-quality photometry or image artefacts.

\subsection{Simulations}\label{sec:baseline-sims}
Our simulations consist of mock \textsc{Healpix} skies that assume a number count modulation strictly from a kinematic dipole, with the exception of the addition of the ecliptic bias correction for CatWISE \citep[][]{Secrest_2021,Secrest2022,10.1093/mnras/stad2322}.

Under this interpretation, the expectation value at each pixel $\mathbf{\hat{n}}_i$ is given by Equation~\ref{eq:expectation-value}. To correct for the ecliptic bias, for CatWISE only we multiply Equation \ref{eq:expectation-value} by
\begin{align}
f_\text{ecl}(\hat{\mathbf{n}}_i)\equiv1-\Upsilon_\text{ecl}c_\text{ecl}|b_\text{ecl}(\hat{\mathbf{n}}_i)|,\label{eq:ecliptic-bias}
\end{align}
where $b_\text{ecl}(\hat{\mathbf{n}}_i)$ is the ecliptic latitude of $\mathbf{\hat{n}}_i$ and $c_\text{ecl}\!=\!9.15\times10^{-4}$ \citep{Secrest2022}.

To evaluate Equations \ref{eq:expectation-value} and \ref{eq:ecliptic-bias} and find the expectation values for the simulation, we draw parameters from the priors
\begin{align*}
\pi(v)&\sim 20\cdot v_\text{CMB}\cdot u\\
\pi(l)&\sim360\cdot u\\
\pi(b)&\sim\sin^{-1}(2\cdot u-1)\\
\pi(\bar{N})&\sim \bar{N}_*(0.2\cdot u+0.9)\\
\pi(\Upsilon_\text{ecl})&\sim4\cdot u-2,
\end{align*}
where $u\in[0,1]$ is a uniform random variable, $\bar{N}_*$ is the mean value per unmasked pixel of the sample the simulation is modelled on, and $\Upsilon_\text{ecl}$ is included only for the CatWISE simulations. \revision{We choose these priors to} match \revision{those} used in \LA, \revision{which allows us to leverage their} \red{$\log R_\text{obs}$} values \revision{and translate them} to $T$ using Equation~\ref{eq:T}. \revision{Here, $\pi(v)$ encloses previous measurements of the dipole \citep[for recent reviews see, e.g.][]{PEEBLES2022169159,2022JHEAp..34...49A,KumarAluri_2023,2025NatRP...7...68S}, $\pi(l)$ and $\pi(b)$ uniformly sample the sphere, $\pi(\bar{N})$ captures up to ten percent variance from the mean source count, and $\pi(\Upsilon_\text{ecl})$ allows up to two times $|\Upsilon_\text{ecl}|$, which we found to be adequate.}

When drawing from the prior, we ensure matched and mismatched pairs of parameter sets as described in Section~\ref{sec:nre-background}. We then translate $v$ to the dipole amplitude for the simulation using
\begin{align}
\mathcal{D}(v)=\tilde{\mathcal{D}}\cdot v/v_\text{CMB},
\end{align}
where $\tilde{\mathcal{D}}$ is the expected dipole amplitude from the sample the simulation is modelled on, given by $v_\text{CMB}/c$ for the CMB \citep[][]{PhysRev.174.2168} and $\tilde{\mathcal{D}}_\text{NVSS}=4.31\times10^{-3}$, $\tilde{\mathcal{D}}_\text{RACS-low}=4.27\times10^{-3}$ and $\tilde{\mathcal{D}}_\text{CatWISE}=7.25\times10^{-3}$ (\LA).

With the expectation values, we generate mock skies by drawing a random value for each pixel. For the NVSS, RACS-low and CatWISE simulations, we draw from a Poisson distribution \red{(Equation~\ref{eq:poisson})} with rate parameter $\lambda_i$. For the \textit{Planck} simulations, we draw from a Gaussian \red{(Equation~\ref{eq:gaussian})} with mean $\lambda_i$ and standard deviation equal to that of the actual \textit{Planck} sample. We then apply the same mask to the simulation as is used by the sample it is modelled on, by setting the masked pixels to zero.


\subsection{Results and Discussion}\label{sec:baseline-results-section}

\begin{figure}
    \includegraphics[width=\columnwidth]{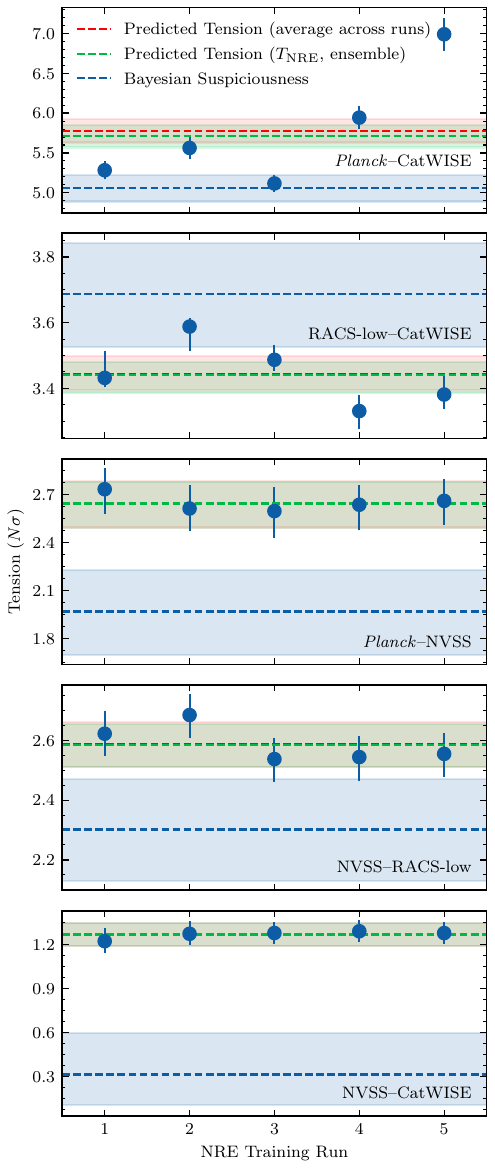}
    \caption{The tension between \textit{Planck}, NVSS, RACS-low and CatWISE using the in concordance \red{$\log R$} distributions learned by our \textsc{tensionnet}. For reference, we show the Bayesian suspiciousness tensions from \LA. \red{The} tension predicted \red{by} each individual NRE \red{across the training runs is represented by the blue points. The predicted (average and ensemble) and Bayesian suspiciousness tensions are represented by the dashed lines, with the shaded regions showing their respective $1\sigma$ errors. In some panels, the average predicted tension is obscured because it aligns with the ensemble result.} The equivalent panel for \textit{Planck}--RACS-low is presented in Figure~\ref{fig:t-comparison}.}
    \label{fig:t-comparison-full}
\end{figure}

\begin{figure}
    \includegraphics[width=\columnwidth]{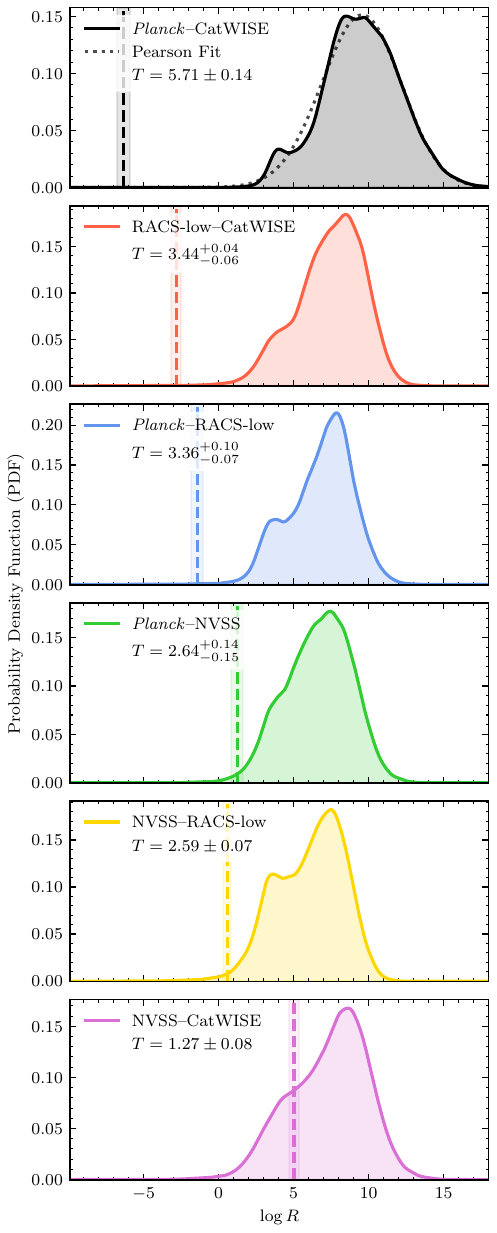}
    \caption{The in concordance \red{$\log R$} distributions learned by our \textsc{tensionnet} for the combinations of \textit{Planck}, NVSS, RACS-low and CatWISE. The dashed line in each panel shows \red{$\log R_\text{obs}$}, which is used to compute $T$, with the shaded region as the $1\sigma$ error. Due to \red{$\log R_\text{obs}$} being far in the tails of the distribution for \textit{Planck}--CatWISE, we instead use a fitted Pearson Type III distribution to compute $T$.}
    \label{fig:distribution-comparison}
\end{figure}

We use our simulations of \textit{Planck}, NVSS, RACS-low, and CatWISE to train NREs for each data set combination and access their in concordance \red{$\log R$} distributions. Given that the \textit{Planck}--RACS-low in concordance \red{$\log R$} distribution captures the ground-truth (recall Figure~\ref{fig:corner}), we use the shape of the distribution to filter-out any excessively broad (peak of the PDF $\ll0.15$) or narrow (peak of the PDF $\gg0.20$) distributions that result when training the individual NREs for the other data set pairs.


We present the tensions predicted by the \textsc{tensionnet} in Figure~\ref{fig:t-comparison-full} and their ensemble in concordance \red{$\log R$} distributions in Figure~\ref{fig:distribution-comparison}. Here, for translating to tension $T$ with Equation~\ref{eq:T}, we use the \red{$\log R_\text{obs}$} values of the data set pairs from \LA. Numerically computing $T$ fails for \textit{Planck}--CatWISE, since \red{$\log R_\text{obs}$} is far in the tails of the empirical distribution. Therefore, we compute $T$ using the analytical CDF of a Pearson Type III distribution fitted using the mean, standard deviation and skewness of the empirical in concordance \red{$\log R$} distribution.

\red{In} Figure~\ref{fig:t-comparison-full}, \red{we see} a clear difference between the tension predicted from the \textsc{tensionnet} and that from Bayesian suspiciousness. \red{As we previously noted when validating our NRE in Section~\ref{sec:nre-validation}, Bayesian suspiciousness is not necessarily expected to align with the output of \red{the} NRE\red{, owing to the different assumptions latent in each of the tension metrics} (see Appendix~\ref{appendix:sec:1} for a comparison of the two). Therefore, we are not concerned about this mismatch.}




\begin{figure}
    \centering
    \includegraphics[width=0.89\columnwidth]{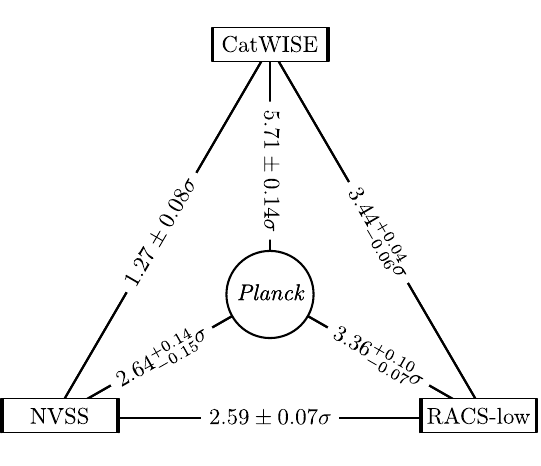}
    \caption{Illustration of the tension between \textit{Planck}, NVSS, RACS-low and CatWISE predicted by our \textsc{tensionnet} under the kinematic interpretation of the dipole, using the ecliptic bias correction for CatWISE. The errors quoted here are $1\sigma$ errors.}
    \label{fig:triangle}
\end{figure}

We find that the relationship between these data sets is virtually unchanged from that found in \LA, which we illustrate in Figure~\ref{fig:triangle}. Here, \textit{Planck} is in severe tension with CatWISE, strong tension with RACS-low, and moderate tension with NVSS\red{. These tensions reinforce} the view that we should be concerned about $\Lambda$CDM \citep[in line with \red{other observational tensions}; for a recent review see][]{2025arXiv250401669D}, the standard kinematic interpretation of the dipole \citep[\red{c.f.}][]{PhysRevD.111.123547,bonnefous2026kinematiccosmicdipoleellis,takeuchi2026generalformulationkinematicdipole}, our understanding of the data, or some combination thereof. CatWISE and NVSS are in no tension, which, given that the surveys have \red{completely independent} systematics \red{(CatWISE is a near-infrared satellite survey and NVSS is a radio-continuum ground-based survey)}, supports the view that \red{they} share a common astrophysical signal \citep[see also][\LA]{Secrest2022}. Finally, RACS-low is in moderate tension with NVSS and strong tension with both \textit{Planck} and CatWISE, \revision{reinforcing the view} that there is an unknown systematic difference in the catalogue itself \revision{(see also \LA)}.

\red{Using} the \textsc{tensionnet}, we have recovered results in line with the literature, \red{reinforcing} the challenge \red{to} $\Lambda$CDM. Whilst this pilot use of the \textsc{tensionnet} uses the standard kinematic interpretation of the dipole, it can be readily extended to use more sophisticated models. Future simulations could model increasingly complex effects, such as over-dispersed source counts \citep[in line with, e.g.][]{6z32-3zf4,10.1093/mnras/stag201}, source detection algorithms, and observational systematics like declination dependence or ionospheric fluctuations. Furthermore, one could use simulations that incorporate the redshift evolution of source populations \citep[see, e.g.][]{Guandalin_2023}, tilted cosmologies \citep[such as those explored in, e.g.][]{martin2025}, and the influence of large scale inhomogeneities \citep[see, e.g.][]{Tiwari_2016}, such as the placement of the observer within a super-void \citep[see, e.g.][]{Rubart2014}. To demonstrate this extensibility, we proceed in Section~\ref{sec:results-intractable} to apply the \textsc{tensionnet} to \revision{CatWISE simulations that \revision{forward-}model Eddington bias}.



\section{Applying the NRE to \revision{CatWISE forward models of Eddington bias}}\label{sec:results-intractable}
In contrast to the CatWISE simulations used in Section~\ref{sec:results-baseline}, we now use `CatSIM' as defined in \OL. Here, we apply the \textsc{tensionnet} to a situation where there is no known likelihood function and there are only synthesised samples under the model. We introduce the simulations below, and then present our results and discuss them.

\subsection{Simulations}\label{sec:simulations-catsim}
While the previous simulations explicitly model the
ecliptic bias as a linear function of source density,
CatSIM proceeds from the assumption that it is ultimately \red{induced}
by Eddington bias.
Namely, where \textit{WISE} has spent less time \red{observing} (lower coverage),
the photometric uncertainty is higher.
In these areas, more sources scatter past the magnitude and colour cut
\red{compared to} regions with \red{higher} coverage and \red{lower} photometric uncertainty. 
\red{Therefore, we observe lower source counts in} areas adjacent to the ecliptic poles\red{, which}
were passed more often by \textit{WISE}\red{.}

To model this process,
instead of drawing Poisson deviates from \textsc{Healpixels},
we produce a forward simulation which maps an initial distribution
of sources to a final density map (for a detailed explanation, see \OL). For completeness, we summarise this process in the steps below, where the only difference here from \OL~is that we model the \citet{Secrest2022} sample \red{rather than} that of \citet{Secrest_2021}, causing a slight difference in magnitude cut and mask:
\begin{enumerate}
    \item We draw sources from a joint W1-W2 distribution
        derived from the CatWISE2020 catalogue, assigning each
        source a position that is sampled uniformly on the sphere.
        The number of sources we draw is simply a parameter to fit,
        which we denote as $N_{\text{init.}}$.
    \item We then use the W1-W2 colour to determine a spectral index $\alpha$
        for each source.
    \item We individually aberrate source positions and boost the magnitudes, introducing the cosmic dipole.
    \item We assign each source a photometric uncertainty depending on 
        the coverage of \textit{WISE} at the associated point on the sky.
    \item Lastly, we apply the mask from \citet{Secrest2022}
        (with an additional $5^\circ$-radius circular mask
        at the north ecliptic pole)
        and only choose sources with
         $9<\text{W1}<16.5$ and $\text{W1} - \text{W2} > 0.8$.
\end{enumerate}

\OL~also introduced the
$\eta_{\text{extra}}$ parameter,
a term that is added in quadrature to the CatSIM photometric errors and
thereby widens their distributions.
Without this parameter, the simulated decrease in source density
at the ecliptic poles is too small compared to the actual CatWISE data,
suggesting that either CatWISE's photometric errors are underestimated
or that an additional systematic, analogous to Eddington bias, is at play.
Accordingly, we use the same parameter in this work,
assuming a Gaussian distribution for the photometric errors
and that $\eta_{\text{extra}}$ applies equivalently for both the
W1 and W2 errors.

We use the same priors for the dipole parameters
as detailed in Section~\ref{sec:baseline-sims}, with the addition of
\revision{\begin{align}
    \pi(N_\text{init.}) &\sim \mathcal{U}_{\text{log}}[3 \times 10^7, 4 \times 10^7],\\
    \pi(\eta_{\text{extra}}) &\sim \mathcal{U}[0, 8],
\end{align}
which, respectively, denote a log uniform prior between 30 million and 40 million sources for the initial CatSIM source count and a uniform prior between zero and eight for the extra photometric error.
\newrevision{Note that most simulated sources will be removed after the colour and magnitude cuts,
leaving a final source count of $\approx 1.6$ million.}
For the $\eta_{\text{extra}}$ prior, as the additional error is added in quadrature, this amounts to scaling the formal photometric uncertainties by at least a factor of one to at most a factor of four.}
Note that,
unlike our simulations from Section~\ref{sec:baseline-sims},
we need not convert an observer speed $v$ to a dipole amplitude $\mathcal{D}$,
since CatSIM directly applies relativistic effects to each source
given \revision{its} simulated magnitude, sky position and spectral index.

\subsection{Results and Discussion}

\begin{figure}
    \includegraphics[width=\columnwidth]{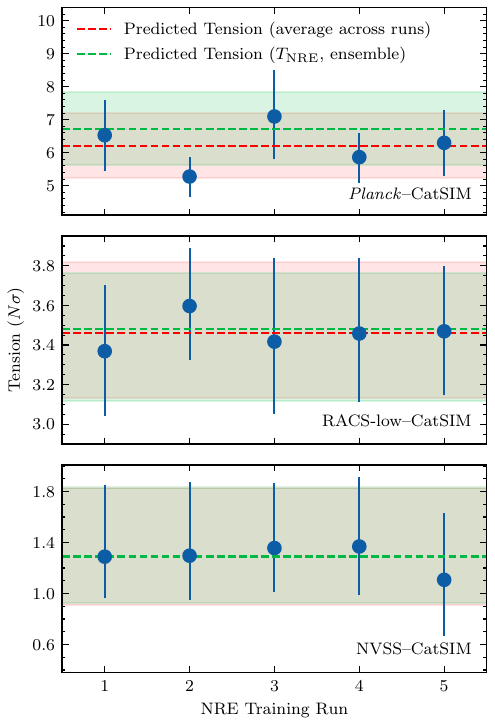}
    \caption{The tension of CatWISE, under the Eddington bias interpretation (CatSIM), with \textit{Planck}, NVSS and RACS-low using the in concordance \red{$\log R$} distributions learned by our \textsc{tensionnet}. \red{The} tension predicted \red{by} each individual NRE \red{across the training runs is represented by the blue points. The average and ensemble predicted tensions are represented by the dashed lines, with the shaded regions showing their respective $1\sigma$ errors. In the bottom panel, the average predicted tension is obscured because it aligns with the ensemble result.} The tensions are consistent with those that use the ecliptic bias correction, as presented in Figure~\ref{fig:t-comparison-full}.}
    \label{fig:t-comparison-full-catsim}
\end{figure}

\begin{figure}
    \includegraphics[width=\columnwidth]{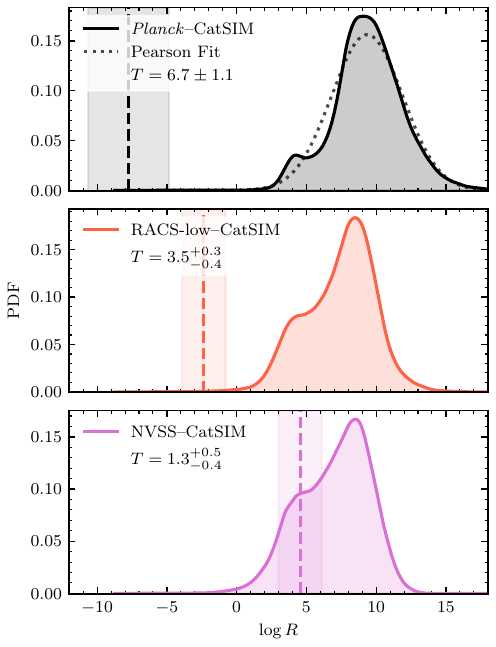}
    \caption{The in concordance \red{$\log R$} distributions learned by our \textsc{tensionnet} for CatWISE, under the Eddington bias interpretation (CatSIM), with \textit{Planck}, NVSS and RACS-low. The dashed line in each panel indicates the $\log R_\text{obs}$ value used to compute $T$, with the shaded region as the $1\sigma$ error. The large errors on \red{$\log R$}, and, by extension, $T$ are due to using an NLE to compute the Bayesian evidence. Similarly to \textit{Planck}--CatWISE, we use a fitted Pearson Type III distribution to compute $T$ for \textit{Planck}--CatSIM. The distributions and \red{tensions} $T$ are consistent with those in Figure~\ref{fig:distribution-comparison}, indicating that the recovered tension is agnostic as to whether one uses the Eddington bias interpretation or the ecliptic bias correction.}
    \label{fig:distribution-comparison-catsim}
\end{figure}

With the Eddington bias interpretation of CatWISE described above, we now train NREs with the CatSIM samples. For translating to tension $T$ using Equation~\ref{eq:T}, we require a measurement of \red{$\log R_\text{obs}$} between the data set pairs under this model. Since the likelihood function required to access \red{$\log R_\text{obs}$} here is intractable, we instead compute it by nested sampling over a Neural Likelihood Estimator (NLE) trained using the same simulation rationale and priors described in Section~\ref{sec:simulations-catsim} (for a detailed outline of the NLE, see \OL). Similarly to \textit{Planck}--CatWISE, for \textit{Planck}--CatSIM we compute $T$ using the analytical CDF of a Pearson Type III distribution that uses the mean, standard deviation and skewness of the original in concordance \red{$\log R$} distribution.

We present the CatSIM tensions and ensemble in concordance \red{$\log R$} distributions predicted by the \textsc{tensionnet} in Figures~\ref{fig:t-comparison-full-catsim} and \ref{fig:distribution-comparison-catsim}, respectively. The large errors on \red{$\log R$} and, by extension, $T$ are due to the greater uncertainty on the Bayesian evidence computed using the NLE (see \OL). Given these measurements, we find that the CatWISE tensions under the Eddington bias interpretation all agree with their counterparts from Section~\ref{sec:baseline-sims} that use the ecliptic bias correction. Our results are in line with \OL, where they find that CatSIM is consistent with the studies that used the ecliptic bias correction \citep[][]{Secrest_2021,Secrest2022,10.1093/mnras/stad2322}, which naturally extends to the same correction used in \LA~and Section~\ref{sec:results-baseline} in this work.

\revision{Whilst the \textit{Planck}--CatSIM and \textit{Planck}--CatWISE tensions are in agreement, we observe a difference in their median value ($6.7\pm1.1\sigma$ compared to $5.71\pm0.14\sigma$), where the tension with \textit{Planck} is greater under the Eddington bias interpretation than under the ecliptic bias correction. This aligns with \OL, where they find that the CatWISE dipole direction is $3\sigma$ away from the CMB expectation under the Eddington bias interpretation, compared to being just under $2\sigma$ away under the ecliptic bias correction \citep[see][]{Secrest_2021}.}

While a significant reduction in tension following the inclusion of a systematic effect would identify it as a likely contributor to the observed discrepancy, our results show that the severe \textit{Planck}--CatWISE tension persists under both interpretations of the ecliptic bias. This indicates that the bias, as currently understood, is not a culprit of tension. This finding highlights the utility of \revision{our} \textsc{tensionnet} \revision{architecture} as a diagnostic tool for future analyses, providing a framework to determine whether specific \revision{effects (such as systematics or astrophysical phenomena)} alleviate or exacerbate the cosmic dipole tension and, by extension, the challenge against $\Lambda$CDM.




\section{Limitations of the NRE}\label{sec:limitations}
The primary hurdle that our \textsc{tensionnet} faces is the need for simulations\red{. In} particular\red{,} the total number of simulations required and the cost to generate them \red{represent significant challenges}.



The computational demand of our simulations varies significantly with the complexity of the underlying model. For our initial simulations that used Poissonian or Gaussian deviates, generating a suite of $900\,000$ samples requires only 1.8 minutes using 48 Intel Xeon Platinum 8268 (2.9~GHz) cores. In contrast, producing the same number of forward-modelled CatSIM samples on comparable hardware (48 Intel Xeon Platinum 8274 cores at 3.2~GHz) takes roughly 140 hours—an increase in runtime of nearly 4\,700 times. \red{Therefore, we suggest that future studies could explore GPU-accelerated methods to reduce such a bottleneck.}

\red{Alternatively, future studies could consider strategies to reduce the number of simulations required for training the NRE.} On this front, since effectively sampling the prior volume takes many simulations when the priors are large, reducing the prior width can help decrease the total number of simulations required. In addition, since the value of $R$ increases with wider priors \citep{Handley2019} and the performance of NREs is known to degrade as the absolute value of the log ratio they are estimating becomes large (\BE), reducing the prior widths could also improve the predictions of the NRE. \red{Whilst} our current setup uses amortised training, where all of the simulations are generated beforehand\red{, future studies could consider} sequential training methods \citep[see, e.g.][]{greenberg2019automaticposteriortransformationlikelihoodfree,https://doi.org/10.5281/zenodo.5043706} \red{such as online learning, where the required simulations are generated at each training epoch.} 



\section{Conclusions}\label{sec:conclusion}
In this work, we presented a \red{novel simulation-based inference (SBI) architecture} \revision{that measures} the cosmic dipole tension\revision{, while being extensible to any systematic or astrophysical effect that can be forward-modelled}. By adapting the \textsc{tensionnet} formalism introduced by \BE, we trained an ensemble of Neural Ratio Estimators (NREs) on \textsc{Healpix} images of the sky to predict the log \revision{Bayesian evidence} ratio \revision{(}$\log R$\revision{)}. Here, we applied simulation-based tension quantification to the cosmic dipole for the first time, measuring the tension between \textit{Planck}, NVSS, RACS-low, and CatWISE.

Our validation tests demonstrate that the \textsc{tensionnet} accurately recovers the ground-truth in concordance \red{$\log R$} distribution, matching the results of nested sampling ($\log\mathscr{D}_{\mathrm{KL}} = -3.99$). Applying the \textsc{tensionnet} to current data sets, we measured a severe \revision{$\approx5.7\sigma$} tension between \textit{Planck} and CatWISE under the kinematic interpretation of the dipole \revision{using the} ecliptic bias correction. In addition, we used the \textsc{tensionnet} to show that NVSS and CatWISE are in no tension, \red{suggesting} that they share a common astrophysical signal \citep[see also][\LA]{Secrest2022}. Conversely, RACS-low being in strong or moderate tension with all of the other surveys \revision{continues to} suggest that it contains an unknown systematic difference (in line with \LA).

A primary advantage of the \textsc{tensionnet} is its ability to incorporate \revision{complex} effects where the likelihood is analytically intractable. We demonstrated this by applying it to CatSIM, a forward simulation of the CatWISE quasar sample \revision{that models the ecliptic bias as an} Eddington bias \revision{with} increased photometric errors. We found that this treatment of the ecliptic bias does not alleviate the discrepancy, yielding a tension of \revision{$\approx6.7\sigma$} with \textit{Planck}. These results reinforce the growing body of evidence \citep[\revision{see,} e.g.][]{Secrest_2021,Secrest2022,Wagenveld_2023,6z32-3zf4} challenging the Cosmological Principle and the standard $\Lambda$CDM model. The persistence of the severe tension suggests that this systematic, as we currently understand it, is not a culprit of tension.

Despite the flexibility of the \textsc{tensionnet}, its application is subject to the practical limitations of \red{the} simulations. Primarily, the computational cost of generating the required number of simulations remains significant; for forward-modelled approaches like CatSIM, the simulation process can be thousands of times more expensive than simple Poissonian realisations, posing a challenge for scaling to even more complex models. Future implementations may benefit from \red{GPU-accelerated methods}, sequential training\red{,} or narrower priors to reduce the total number of simulations required.

Ultimately, the future of dipole analyses will rely on our ability to model even more complex \revision{effects}, such as \revision{those arising from} local voids, \revision{tilted cosmologies,} redshift evolution, and \revision{instrumental systematics.} \red{Our \textsc{tensionnet} architecture} provides the necessary machinery to test these models as upcoming facilities--including the Vera C. Rubin Observatory Legacy Survey of Space and Time \citep[LSST;][]{Ivezic_2019}, \textit{Euclid} \citep{2024arXiv240513491E}, and the Square Kilometre Array \citep[SKA;][]{Bacon2020}--begin to provide unprecedented source counts. By moving beyond the limitations of analytical likelihoods, we can ensure that our tension quantification remains robust as we enter the era of sub-percent precision cosmology.

\section*{Acknowledgements}
\red{We thank Will Handley for useful discussions on comparing Bayesian suspiciousness to the tension derived from in concordance $\log R$ distributions.} \revision{We also thank the anonymous referee for their helpful comments, which greatly improved this manuscript. Moreover, we} thank Matthew Sale for providing help and expertise with the high performance computing used in this work \revision{and} Jordan Winstanley for helpful discussions on high performance computing. \revision{In addition, we thank Daniel J. Ballard for useful comments on this manuscript.} \red{We also extend our gratitude to Stefan W. Duchesne, Emil Lenc and Tara Murphy for valuable discussions regarding the RACS catalogues. MLS acknowledges support from the Astronomical Society of Australia Student Travel Assistance Scheme, the CSIRO Space and Astronomy Postgraduate Student Program, and the MaxEnt and Bayesian Association of Australia Student Sponsorship. HTJB acknowledges support from the Kavli Institute for
Cosmology Cambridge and the Kavli Foundation.} MLS is supported by the Australian Government Research Training Program (RTP) Scholarship. OTO is supported by the University of Sydney Postgraduate Award.

This work made use of the \textit{Planck} Public Data Release 3 \citep{2020ipac.data.I558P}, the \textsc{BeyondPlanck} Data Release II \citep{2023A&A...675A...1B}, the National Radio Astronomy Observatory Very Large Array Sky Survey \citep{Condon_1998}, the Rapid Australian Square Kilometre Array Pathfinder Continuum Survey \citep{McConnell_2020}, the CatWISE2020 data release \citep{Eisenhardt_2020}, the Two Micron All Sky Survey Redshift Survey \citep{Huchra_2012}, and the NASA/IPAC Extragalactic Database.

This scientific work uses data obtained from Inyarrimanha Ilgari Bundara, the CSIRO Murchison Radio-astronomy Observatory. We acknowledge the Wajarri Yamaji People as the Traditional Owners and native title holders of the Observatory site. CSIRO’s ASKAP radio telescope is part of the Australia Telescope National Facility (\url{https://ror.org/05qajvd42}). Operation of ASKAP is funded by the Australian Government with support from the National Collaborative Research Infrastructure Strategy. ASKAP uses the resources of the Pawsey Supercomputing Research Centre. Establishment of ASKAP, Inyarrimanha Ilgari Bundara, the CSIRO Murchison Radio-astronomy Observatory and the Pawsey Supercomputing Research Centre are initiatives of the Australian Government, with support from the Government of Western Australia and the Science and Industry Endowment Fund.

This research was undertaken with the assistance of resources from the National Computational Infrastructure (NCI Australia), an NCRIS enabled capability supported by the Australian Government and the Sydney Informatics Hub, a Core Research Facility of the University of Sydney.

This work used the \textsc{Python} packages \textsc{Anesthetic} \citep{anesthetic}, \textsc{Astropy} \citep{2022ApJ...935..167A}, \textsc{Healpy} \citep{2005ApJ...622..759G,Zonca2019}, \textsc{Matplotlib} \citep{4160265}, \textsc{NNhealpix} \citep{Krachmalnicoff2019}, \textsc{Numpy} \citep{Harris_2020}, \red{\textsc{Optuna} \citep{akiba2019optuna},} \textsc{Scipy} \citep{2020NatMe..17..261V}, \textsc{TensorFlow} \citep{tensorflow2015-whitepaper}, and \textsc{Ultranest} \citep{2021JOSS....6.3001B}.


\section*{Data Availability}
The ASKAP data used in this work is publicly available from the CSIRO
ASKAP Science Data Archive (CASDA; \url{https://research.csiro.au/casda}) under the project code AS110. Any other data used in this work are publicly available online. Any samples used in this paper will be made available upon reasonable request to the authors.
 



\bibliographystyle{mnras}
\bibliography{bibliography} 




\appendix

\section{\revision{Derivation of the NRE output}}\label{appendix:derivation}
\revision{As introduced in Section \ref{sec:nre-background}, if trained on an equal number of data set pairs ($A$ and $B$) drawn from their joint distribution $P(A,B)$ and the product of their marginal distributions $P(A)P(B)$, the output of the NRE tends towards $\log r = \log R$. Here, we detail the derivation of this result (see also section 4 in \BE).}

\revision{We first treat the output of the NRE as a function $f(A,B)$, where the network is trained as a binary classifier used to distinguish
samples of the joint distribution from samples of the product of the marginals.
We draw training data from $P(A,B)$ with probability $p$ and label $y_i=1$, and also
from $P(A)P(B)$ with probability $1-p$ and label $y_i=0$. A sigmoid activation maps
the network output onto $(0,1)$, so that it can be read as a class probability,}
\begin{align}
    \tilde{f}(A,B)\equiv S_\sigma\bigl(f(A,B)\bigr)=\frac{e^{f(A,B)}}{1+e^{f(A,B)}},
    \label{eq:nre-sigmoid}
\end{align}
\revision{and the network is trained by minimising the binary cross entropy loss function}
\begin{align}
    l=-\frac{1}{N}\sum^N_{i}\Bigl[y_i\log\tilde{f}(A_i,B_i)
      +(1-y_i)\log\bigl(1-\tilde{f}(A_i,B_i)\bigr)\Bigr].
    \label{eq:loss-derivation}
\end{align}

\revision{In the limit of infinite training data, $N\to\infty$, the empirical average in
Equation~\ref{eq:loss-derivation} converges to an expectation over the two populations,}
\begin{align}
    l=-\int \Bigl[&\,p\,P(A,B)\log\tilde{f}(A,B)\nonumber\\
    &+(1-p)P(A)P(B)\log\bigl(1-\tilde{f}(A,B)\bigr)\Bigr]\,\mathrm{d}A\,\mathrm{d}B.
    \label{eq:loss-continuous}
\end{align}
\revision{Therefore, given that the loss function is minimised, the functional derivative becomes zero,}
\begin{align}
    \frac{\delta l}{\delta \tilde{f}}
    =-\frac{p\,P(A,B)}{\tilde{f}(A,B)}
     +\frac{(1-p)P(A)P(B)}{1-\tilde{f}(A,B)}=0,
    \label{eq:nre-convergence}
\end{align}
\revision{and rearranging gives}
\begin{align}
    \tilde{f}(A,B)&=\frac{\frac{pP(A,B)}{(1-p)P(A)P(B)}}{1+\frac{pP(A,B)}{(1-p)P(A)P(B)}},
\end{align}
\revision{where the numerator is precisely $e^{f(A,B)}$. Taking the logarithm therefore recovers the network output directly,}
\begin{align}
    f(A,B)=\log\frac{p\,P(A,B)}{(1-p)P(A)P(B)}.
\end{align}
\revision{For balanced classes, $p=1/2$, the prefactor vanishes and the optimally trained
NRE learns the log-ratio,}
\begin{align}
    f(A,B)=\log\frac{P(A,B)}{P(A)P(B)}=\log r=\log R.
\end{align}

\section{\red{Hyperparameter optimisation}}\label{appendix:sec:2}
\red{To infer the optimal hyperparemters for our NRE architecture, we used \textsc{Optuna} \citep{akiba2019optuna}. Specifically, we conducted an initial search over the dropout rate and the number of filters, neurons, and layers. Here, we trained 43 NREs for a maximum of 15 epochs following the same rationale as in Section~\ref{sec:training-nre}, except with 200\,000 simulated data set pairs (following the same split percentages) and a patience of 10. We gauged the performance of each individual NRE by comparing its in concordance $\log R$ distribution to the nested sampling ground-truth from Section~\ref{sec:nre-validation}, monitoring the Kullback-Leibler Divergence (see Equation~\ref{eq:kl}). We found that the initial search heavily favoured one filter, so we removed that hyperparameter from the search.}

\red{We then conducted a second search over the dropout rate, the initial learning rate, the activation function (\textsc{ReLU}, \textsc{LeakyReLU}, \textsc{GELU}, and \textsc{ELU}), and the number of neurons and layers. Here, we used one filter and trained 369 NREs following the same rationale as above, except with 100\,000 simulated data set pairs and a variable learning rate (since it is part of the hyperparameter search). Based on the \textsc{Optuna} studies, we adopted the architecture and training procedure described in Section~\ref{sec:nre-architecture}.}

\section{\red{Comparing NREs to} Bayesian Suspiciousness}\label{appendix:sec:1}
To investigate the difference between the tensions predicted by the NRE and those measured using Bayesian suspiciousness, we construct a toy problem with an analytically tractable tension and in concordance \red{$\log R$} distribution. 

\subsection{Bayesian Suspiciousness}
We first summarise Bayesian suspiciousness, which was introduced by \cite{Handley2019} as a prior invariant measure of tension and concordance. As we mentioned in Section~\ref{sec:Rstat}, they directly show in Equation~\ref{eq:Rpriordependence} that $R$ is prior dependent, whereby wider priors inflate the value of $R$. To remove this prior dependence, they then introduce the information ratio
\begin{align}
    \log I=\mathscr{D}_A+\mathscr{D}_B-\mathscr{D}_{AB},
\end{align}
whereby $I$ increases similarly to $R$ when priors are widened. Here, $\mathscr{D}_D$ is the Kullback-Leibler Divergence \citep{10.1214/aoms/1177729694}
\begin{align}
    \mathscr{D}_D=\int\mathcal{P}_D(\theta)\log \frac{\mathcal{P}_D(\theta)}{\pi(\theta)}~d\theta=\left< \log\frac{\mathcal{P}_D}{\pi} \right>_{\mathcal{P}_D},
\end{align}
which quantifies the information gained going from the prior to the posterior for data set $D$, where $D$ can be either of our individual data sets $A$ or $B$, or our joint data set $AB$.

Therefore, subtracting $\log I$ from $\log R$ yields a prior invariant quantity, Bayesian suspiciousness
\begin{align}
    \log S=\log R - \log I.\label{eq:bayesian-suspiciousness}
\end{align}
Similarly to $R$, if $S\ll1$ the data sets are in tension and if $S\gg1$ the data sets are in suspiciousness concordance.

Under the assumption of Gaussian posteriors, Bayesian suspiciousness can be converted to $N\sigma$ tension, since the quantity $d-2\log S$ follows a $\chi^2_d$ distribution. Here, $d$ is the effective number of parameters constrained by both of the data sets
\begin{align}
    d=\tilde{d}_A+\tilde{d}_B-\tilde{d}_{AB},
\end{align}
and $\tilde{d}_D$ is the Bayesian model dimensionality \citep{PhysRevD.100.023512} of data set $D$
\begin{align}
    \tilde{d}_D=2\left<\left(\log\frac{\mathcal{P}_D}{\pi}\right)^2\right>_{\mathcal{P}_D}-~~~2\left<\log\frac{\mathcal{P}_D}{\pi}\right>^2_{\mathcal{P}_D}.
\end{align}
The tension probability $p$-value of the observation can thus be easily obtained as
\begin{align}
    p_T&=\int^\infty_{d-2\log S}\chi^2_d(x)~dx\\
    &=\int^\infty_{d-2\log S}\frac{x^{d/2-1}e^{-x/2}}{2^{d/2}\Gamma(d/2)}~dx=\frac{\Gamma\left(\frac{d}{2},\frac{d-2\log S}{2}\right)}{\Gamma(d/2)},
\end{align}
where $\Gamma(s)$ and $\Gamma(s,x)$ are the regular and upper incomplete gamma functions. Using the inverse error function, the reported tension is thus
\begin{align}
    N_\sigma=\sqrt{2}~\text{\red{e}rf}^{-1}(1-p_T).
\end{align}

\subsection{Toy Problem Analysis}
We use \red{an} analytically tractable toy problem \red{similar to that} defined in section \revision{6} of \BE. \red{To evaluate all of the expressions we use the package \textsc{lsbi}\footnote{\url{https://github.com/handley-lab/lsbi}}.} We first define our prior and likelihoods to be Gaussian
\begin{align}
    \mathcal{L}(D|\theta)&=\mathcal{N}(M\theta+m,C),\\
    \pi(\theta)&=\mathcal{N}(\mu,\Sigma),
\end{align}
and for our observed data sets to use a linear model
\begin{align}
    D = M\theta+m\pm \sqrt{C}.
\end{align}
Here, the data model (experiment) is defined by $M$ and $m$, $\theta$ are the model parameters, $C$ is the covariance of the likelihood, $\mu$ and $\Sigma$ are the mean and covariance of the prior, and \red{$\mathcal{N}(a,b)$ is a Gaussian PDF with mean $a$ and covariance $b$.} Each experiment uses a different $M$, $m$ and $C$ to observe the same parameters $\theta$. Here, the Bayesian evidence for each experiment is analytically tractable and is
\begin{align}
    \mathcal{Z} = \mathcal{N}(m+M\mu,C+M\Sigma M^T).
\end{align} 

For our experiments, we use $d=50$ data points with $n=3$ dimensions. We define $M$ to be a $d\times n$ matrix of uniform random numbers between 0 and 1, $m$ and $\mu$ to be vectors of uniform random numbers between 0 and 1, and $C$ to be $0.01\mathcal{I}$, where $\mathcal{I}$ is the identity matrix. Even though $M$ and $m$ are unique to each experiment, they use the same prior with $\Sigma$ and $\mu$. For our comparison, we employ three prior widths, namely $\Sigma=0.1\mathcal{I}$, $1\mathcal{I}$, and $100\mathcal{I}$.

\red{For six approximately linearly spaced tensions between $0.5\sigma$ and $3\sigma$, we} draw in concordance (matched) realisations of our experiments from the joint distribution $\mathcal{Z}_{AB}=P(D_A,D_B)$, \red{which we call the reference observation pairs}. We then evaluate the analytical in concordance \red{$\log R$} distribution between the experiments and their analytical \red{$\log R_\text{obs}$} values. With these in hand, we translate their \red{$\log R_\text{obs}$} to tension $T$ using Equation~\ref{eq:T}, and evaluate their Bayesian suspiciousness $N\sigma$ tension.

With the ground-truth tension and Bayesian suspiciousness tension outlined above, all that is left is using the NRE to predict the tension. For each prior width $\Sigma$ and \red{reference} observation \red{pair} \red{(each of which are} at a different level of tension\red{)}, we generate $1\,000\,000$ pairs of observations of experiments $A$ and $B$ that are equally mixed between matched and mismatched observations. We use 80 percent of them for training and 20 percent for validation. We use the NRE schematic from figure 1 in \BE, with five dense layers each with 25 nodes, using \textsc{ReLU} activations. In addition, we use an ADAM optimiser with a learning rate of $10^{-4}$. For each $\Sigma$ and \red{reference} observation \red{pair}, we train five NREs for a maximum of 1000 epochs with a batch size of 1000 and use early stopping with a patience of 20, monitoring validation loss. With the trained NREs, we predict the empirical in concordance \red{$\log R$} distribution from a new set of 5000 in concordance observations of our experiments. Finally, we compute the average NRE tension $T$ \red{for each reference observation pair} by passing the analytical \red{$\log R_\text{obs}$} through the empirical in concordance \red{$\log R$} distributions.

\begin{figure}
    \includegraphics[width=\columnwidth]{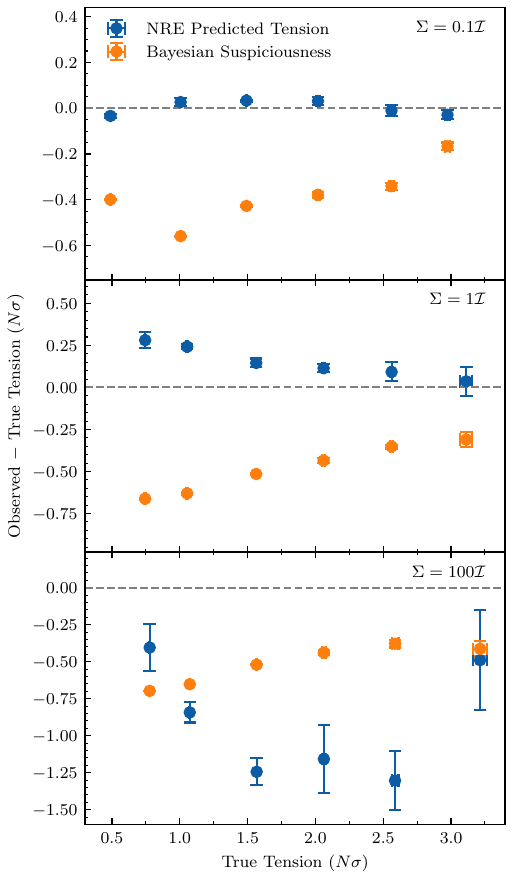}
    \caption{Comparison between the recovered tensions of a toy problem, using a) the in concordance \red{$\log R$} distribution learned by an NRE and b) Bayesian suspiciousness. Each panel uses a different prior width, with the smallest prior of $\Sigma=0.1\mathcal{I}$ in the top panel and the largest prior of $\Sigma=100\mathcal{I}$ in the bottom panel. \red{Each panel uses the same set of six reference observation pairs, which are each in different amounts of true (analytical) tension.} We find that Bayesian suspiciousness consistently underestimates tension and that the predictions from the NRE align better with the analytical truth (except for $\Sigma=100\mathcal{I}$, where the performance of NREs is known to degrade with wide priors, since the absolute value of the log ratio they are predicting is large; see \BE). Since our \textsc{tensionnet} operates in the regime between $\Sigma=0.1\mathcal{I}$ and $1\mathcal{I}$, we assert that its predictions align with the ground-truth.}
    \label{fig:toy-problem}
\end{figure}

We present our results in Figure~\ref{fig:toy-problem}, where Bayesian suspiciousness consistently underestimates the tension compared to the ground-truth. Here, for $\Sigma=0.1\mathcal{I}$ and $1\mathcal{I}$, the predictions of the NRE align better with the ground-truth. At $\Sigma=100\mathcal{I}$, the extended width of the priors causes the absolute value of $\log R$ to increase, where the performance of the NRE is known to degrade (\BE). From comparing our in concordance \red{$\log R$} distributions in Figures \ref{fig:corner}, \ref{fig:distribution-comparison} and \ref{fig:distribution-comparison-catsim} to figure 4 in \BE, it is evident that our \textsc{tensionnet} is operating in the regime between $\Sigma=0.1\mathcal{I}$ and $1\mathcal{I}$. Therefore, we assert that the tensions predicted by our \textsc{tensionnet} reflect the ground-truth.

\bsp	
\label{lastpage}
\end{document}